\documentclass[a4paper,11pt]{article}

\usepackage{jcappub} 
\usepackage{amsfonts}
\usepackage{amsmath}
\usepackage{amssymb}
\usepackage{mathrsfs}
\usepackage{subfigure}
\usepackage{graphicx}
\usepackage{epstopdf}
\usepackage{float}
\usepackage{color}
\usepackage{bm}
\usepackage{ulem}
\usepackage{xcolor}
\usepackage{appendix}
\usepackage{booktabs}
\usepackage{hyperref}
\usepackage{array}
\usepackage{cancel}

\usepackage[T1]{fontenc} 

\title{\boldmath Gravitational Lensing of Spherically Symmetric Black Holes in Dark Matter Halos}

\author[a]{Yi-Gao Liu,}
\author[b]{Chen-Kai Qiao,}
\author[a]{Jun Tao}

\affiliation[a]{College of Physics, Sichuan University, Chengdu, 610065, China}
\affiliation[b]{College of Science, Chongqing University of Technology, Chongqing, 400054, China}

\emailAdd{liuyigao@stu.scu.edu.cn}
\emailAdd{chenkaiqiao@cqut.edu.cn}
\emailAdd{taojun@scu.edu.cn}

\abstract{
	The gravitational lensing of supermassive black holes surrounded by dark matter halo has attracted a great number of interests in recent years. However, many studies employed simplified dark matter density models, which makes it very hard to give a precise prediction on the dark matter effects in real astrophysical galaxies. In this work, to more accurately describe the distribution of dark matter in real astrophysical galaxies, we study the gravitational lensing of black holes in astrophysical dark matter halo models (Beta, Burkert, Brownstein, and Moore). The deflection angle is obtained using a generalized Gibbons-Werner approach. The visual angular positions and the Einstein rings are also calculated by adopting the gravitational lens equation. Specifically, we choose the supermassive black holes in Milky Way Galaxy, Andromeda galaxy (M31), Virgo galaxy (M87), and ESO138-G014 galaxy as examples, including the corresponding fitted value of dark matter halos. The results suggest that the dark matter halo described by the Beta model has non-negligible influences on the gravitational deflection angle and gravitational lensing observations. However, the Burkert, Brownstein, and Moore models have relatively small influences on angular position of images and the Einstein ring.
}

\begin{document}
	\maketitle
	\flushbottom

\section{Introduction}

In the past years, abundant astrophysical observations revealed that our universe is dominated by dark matter and dark energy. The cosmic microwave background observation shows that 26.8\% of our universe has consisted of dark matter, and 68.3\% of our universe is made up of dark energy, while only 4.9\% of universe is composed of baryonic matter \cite{Planck:2013oqw, Planck:2013pxb, WMAP:2010sfg, Planck:2018vyg}. The dark matter has non-negligible influences on the galactic and cosmological predictions. Especially, the large-scale structures formation of the universe \cite{Blumenthal:1984bp, Davis:1985rj, Reddick:2012qy, Rocha:2012jg}, bullet clusters \cite{Markevitch:2003at, Clowe:2006eq}, star motions in galaxies (galactic rotational curves \cite{Rubin:1970zza, Sofue:2000jx, Bertone:2016nfn}), and gravitational lensing can hardly get desired explanations without assuming the dark matter. Furthermore, in past few years, a number of X-ray and matter-antimatter satellites and detectors (such as Chandra, Fermi-LAT, AMS and DAMPE) reported anomalous excess of X-rays and other cosmic-rays \cite{Bertone:2004pz, Fermi-LAT:2017opo, DAMPE:2017fbg, AMS:2021nhj}. The dark matter near the galaxy center provides a viable explanation on such excess. There are a number of dark matter candidates proposed in the past few decades, and most of them were given by some unknown non-baryonic particles generated in theories beyond the Standard Model \cite{Jungman:1995df, Bertone:2004pz}. In recent years, the weakly-interactive massive particles (WIMPs) and axions have attracted considerable interest in dark matter detection experiments \cite{Nomura:2008ru, AMS:2016oqu,PandaX-II:2016vec, XENON:2018voc, Schumann:2019eaa, ADMX:2019uok, Chadha-Day:2021szb}.

The dark matter vastly influences the behavior of galaxies / galaxy clusters, for it could produce extremely strong gravitational potential, heavily affecting the particles and stars motions in galaxies / galaxy clusters. The dark matter usually forms halo structures in the galaxy and galaxy clusters \cite{Wang:2019ftp, Zavala:2019gpq, Bhattacharya:2011vr}. The range and mass of dark matter halos are very large \cite{Bhattacharya:2011vr}. The supermassive black hole in a galaxy center is usually surrounded by a dark matter halo with a typical length scale $h \sim 10$ kpc. The dark matter halo structure strongly affects the galactic rotational curves \cite{Rubin:1970zza, Sofue:2000jx, Bertone:2016nfn, Corbelli:1999af, Corbelli:2003sn, Sofue:2020rnl}, the motions of matter in Bullet cluster collision observations \cite{Clowe:2006eq}, the gravitational lensing of massive objects \cite{Karamazov:2021hwa, Karamazov:2021itg, QiQi:2023nex}, the particle motions and chaos \cite{Zhou:2022eft, Wang:2019ftp}, the black hole shadow and its polarized image \cite{Jusufi:2020cpn, Jusufi:2019nrn, Vagnozzi:2022moj, Das:2021otl}. Through simulation studies and astrophysical observations, many dark matter halo models have been proposed, and most of them suggest that the dark matter distribution is spherically symmetric. The most popular halo models include the Navarro-Frenk-White (NFW) model \cite{Navarro:1995iw, Navarro:1996gj}, Einasto model \cite{Dutton:2014xda, Graham:2005xx, Navarro:2008kc}, Beta model \cite{Navarro:1994hi, Cavaliere:1976tx}, Burkert model \cite{Burkert:1995yz, Salucci:2000ps}, Brownstein model \cite{Brownstein:2009gz}, and Moore model \cite{Moore:1999gc}. These models have become extremely valuable in phenomenological studies, which restrain the indirect and direct dark matter detections \cite{Iocco:2015xga, Pato:2015tja, Cuoco:2016eej, Salucci:2018hqu}. 

The gravitational lensing is one of the most significant methods to extract knowledge and properties from massive objects in the galaxy and our universe, especially in the presence of dark matter. The enormous amount of dark matter and luminous matter in galaxies and galaxy clusters makes the light rays deflected and distorted during the propagation, producing single or multiple images for distant light sources. Essentially, the gravitational lensing is nothing but relevant to the particle motions. It is natural to infer that dark matter, which produce strong gravitational potential, may have notable influences on gravitational lensing observations in the galactic scales. The anomalous excess of X-rays and other cosmic-rays reported by Chandra, Fermi-LAT, AMS and DAMPE \cite{Bertone:2004pz, Fermi-LAT:2017opo, DAMPE:2017fbg, AMS:2021nhj} indicates the large amount dark matter near the galaxy center around the supermassive black holes (which may generate potentially observable effects on the gravitational lensing). In the last few decades, gravitational lensings have often been used to constrain non-luminous matters, such as dark matter distributions, and other massive compact objects \cite{Zumalacarregui:2017qqd, Delos:2023fpm, Despali:2023nel}. In astrophysical observations, the gravitational deflection angle of light, gravitational lens equation, and Einstein ring are crucial quantities, closely connected with the distributions and properties of gravitational sources. These have been extensively researched in many gravitational systems \cite{Virbhadra:2002ju, Virbhadra:2007kw, Virbhadra:2008ws, Tsukamoto:2012xs, Cunha:2018acu, Gralla:2019xty, Zhang:2021ygh, Qiao:2021trw, Islam:2020xmy, Fu:2022yrs, Guo:2022muy, Virbhadra:2022iiy, Jha:2022vun, Kuang:2022xjp, Kuang:2022ojj, Belhaj:2022nuk, Ghosh:2022mka, AbhishekChowdhuri:2023ekr, Tsukamoto:2022uoz, Soares:2023err, Nash:2023zza, Soares:2023uup, Hu:2023mai}.

Based on the aforementioned reasons, it is extremely important to study the dark matter effects on gravitational lensing observations in galaxies, especially for the gravitational lensing of supermassive black hole in the galaxy center. Recently, the gravitational lensing of black holes surrounded by dark matter has been extensively studied in literature \cite{Jusufi:2019nrn, Ovgun:2018oxk, Pantig:2020odu, Atamurotov:2021hck, Qiao:2022nic, Gao:2023ltr, Molla:2023yxn, Pantig:2021zqe, Pantig:2022whj, Pantig:2022toh, DeLuca:2023laa, Datta:2023zmd}. These studies suggested that dark matter has a significant effect on gravitational deflection and gravitational lensing. However, many studies employed simplified dark matter density models \cite{Pantig:2020odu, Pantig:2021zqe, Atamurotov:2021hck, Qiao:2022nic, Gao:2023ltr}, from which it is hard to give a precise dark matter mass distribution (as well as its influences on central black holes) in real galaxies. Therefore, using precise and authentic dark matter models rather than over-simplifying ones in the gravitational lensing of central black holes is necessary. Until very recently,  \'Ovg\"un et al first considered the gravitational lensing of black hole within the cold dark matter halo model with an NFW density profile \cite{Pantig:2022toh}, where a real galactic dark matter distribution goes beyond the over-simplified model is included in analytical studies.

Inspired by \'Ovg\"un's work, we are interested in the gravitational lensing of black holes interplayed with dark matter medium in galaxy centres. Using several astrophysical dark matter halo models (Beta, Burkert, Brownstein, and Moore), we are committed to give a precise description on the galactic dark matter halo. In this work, we study the gravitational deflection of spherically symmetric black holes in dark matter halos. The dark matter distributions in galaxy and halo structures are given by several phenomenological dark matter density profiles: the Beta, Burket, Brownstein, and Moore models. Furthermore, using a generalized Gibbons and Werner (GW) approach \cite{Gibbons:2008rj, Werner:2012rc, Huang:2022gon, Huang:2022soh, Huang:2023bto}, the gravitational deflection angles of light are derived and calculated. Additionally, we also study the dark matter halo effects on gravitational lensing observations. The important observables in gravitational lensing are the positions of luminous sources and the lensed images, and they are constrained by gravitational lens equations \cite{Virbhadra:1999nm, Dabrowski:1998ac, Frittelli:1998hr, Perlick:2003vg, Bozza:2006sn, Bozza:2008ev}. One of the most important observables in gravitational lensing, the Einstein ring, is mainly focused in the present work. To directly connect with the astrophysical gravitational lensing observations, it would be helpful to select some real supermassive black holes and dark matter halos in galaxies and galaxy clusters to carry out calculations. We consider the supermassive black holes in Milky Way Galaxy, Andromeda galaxy (M31), Virgo galaxy (M87), and ESO138-G014 galaxy, and the fitted value of dark matter halos in these galaxies are taken into calculations.

This paper is organized as follows. In section \ref{theory}, we derived the effective spacetime metrics for spherically symmetric black holes in dark matter halos. The theoretical treatment (the generalized GW method to gravitational deflection angle, the gravitational lens equation to angular positions of images and Einstein ring) is introduced in section \ref{gravitational lensing}. Section \ref{gravitational deflection angle} gives the analytical results of the gravitational deflection angle for black holes surrounded by dark matter halos. We derive the gravitation deflection angle for both the receiver and the light source at infinity. In section \ref{ring}, the angular positions of lensed images and Einstein ring are investigated by solving the gravitational lens equation. We conclude our works and give perspectives in section \ref{concl}. Appendix \ref{appendixA} compares the effects of dark matter and luminous matter on gravitational deflection angles. Throughout the paper, we adopt the convention $G = c = 1$.

\section{The Spherically Symmetric Black Hole Surrounded by a Dark Matter Halo \label{theory}}

In this section, we give the descriptions of the spacetime metric for black holes surrounded by a dark matter halo. The dark matter distributions in our galaxies and other spiral galaxies are usually modeled by several astrophysical phenomenological models, such as NFW, Einasto, Beta, Burkert, Brownstein, and Moore models \cite{Brownstein:2009gz, Burkert:1995yz, Fukushige:2003xc, Moore:1999gc, Navarro:1994hi, Navarro:1996gj, Salucci:2000ps, Sofue:2020rnl, Cavaliere:1976tx}. For most galaxies, the dark matter halo can be effectively described by a spherically symmetric distribution. For simplicity, we shall focus on the non-rotating black holes surrounded by dark matter halos whose spacetime metric can be expressed by, 
\begin{equation}
    \text{d\textit{s}}^2=-{f}(r) \text{dt}^2 +{f}(r) ^{-1}\text{dr}^2 +r^2\text{d$\theta $}^2+r^2\mathcal{\text{sin}}^2\theta \text{d$\phi $}^2.
\end{equation}
The effects of dark matter halo in gravitational lensing are through its mass profile and mass density. The dark matter halo mass profile is defined as
\begin{equation}
    M_{DM}(r) =4 \pi \int_{0}^{r} \mathcal{\rho}(r') r'^2 dr',
\end{equation}
where ${\rho}(r)$ is the density of dark matter distributions, and the detailed expressions are given by astrophysical phenomenological models. Using the mass profile, the tangential velocity of the test particle in dark matter halo is calculated easily by ${v_{tg}}^2(r) = M(r)/r$. For a test particle in spherical symmetric spacetime, its rotation velocity in the equatorial plane is determined by the metric function $f(r)$ as \cite{Matos:2000ki}
\begin{align}
    {v_{tg}}^2(r)=\frac{r}{\sqrt{f (r)}}\cdot  \frac{d \sqrt{f (r)}}{dr}=\frac{r (\text{dln} \sqrt{{f} (r)})}{dr}.\label{0.1}
\end{align}
According to the rotational velocity in Eq. ({\ref{0.1}}), the metric function of the dark matter halo can be derived by solving the ordinary differential equation Eq. (\ref{0.1})
\begin{align}
   f_{DM}(r)=\mathrm{exp}[2\int_{}^{} \frac{{v_{tg}}^2(r)}{r} \mathrm{d}r].\label{0.2}
\end{align}

In this work, we use several famous spherically symmetric dark matter profiles (Beta, Burkert, Brownstein, Moore). These models were proposed and developed for many years. Cavaliere and Fusco-Femiano proposed the isothermal Beta model in 1976 when studying the distribution of matter in galaxies \cite{Cavaliere:1976tx}. Burkert proposed an empirical profile that successfully fitted the halo rotation curves of four dark matter-dominated dwarf galaxies in 1995 \cite{Burkert:1995yz,  Salucci:2000ps}. In 1999, Moore et al. simulated that the profile has a cusp proportional to $r^{-1.5}$ in both galaxy-sized and cluster-sized halos \cite{Moore:1999gc}. Brownstein showed that the core-modified profile with a constant central density fits excellently well to the rotation curves of both high- and low-surface brightness galaxies in 2009 \cite{Brownstein:2009gz}. These dark matter profiles have been widely used in theoretical predictions and numerical simulations in physics and astronomy. The dark matter distributions of the Beta, Burkert, Brownstein, and Moore models can be expressed as
\begin{subequations}
    \begin{eqnarray}
        &&\rho _{Beta}(x)=\frac{\rho {_0} _{Beta} }{(1+x^{2})^{3/2}},\\
        &&\rho _{Bur}(x)=\frac{\rho {_0}_{Bur} }{(1+x)(1+x^2)},\\ 
        &&\rho _{Bro}(x)=\frac{\rho {_0}_{Bro} }{1+x^3},\\
        &&\rho _{Moo}(x)=\frac{\rho {_0}_{Moo} }{x^{3/2}(1+x^{3/2})},
    \end{eqnarray}
\end{subequations}
where $x=r/h$, $\rho _0$ and $h$ are the characteristic density and radius of dark matter halos respectively. Among these dark matter distribution, the Beta, Burkert, and Brownstein models are cored halo models with a smooth density profile near the galaxy center, and the Moore model is a cusp halo model with a rapidly increased density profile near the galaxy center \cite{Sofue:2020rnl}.

To get the spacetime metric of black holes surrounded by dark matter halos, one can combine the dark matter density and the central black hole mass $M$ to the energy-momentum tensor $T_{\mu\nu} $ in the Einstein field equation. It turns out that the spacetime metric can be decomposed into $f(r)=f_{DM}(r)-\frac{2M}{r}$, where $f_{DM}(r)$ in Eq. (\ref{0.2}) describes the dark matter halo and $-\frac{2M}{r}$ describes the effects of supermassive black hole in the galaxy center, seeing Ref. \cite{Xu:2018wow} for details. Eventually, we get the metric function of the black hole in each dark matter halo model:
\begin{subequations}
    \begin{eqnarray}
        &&f_{Beta}  (r)=e^{-\frac{8 \pi  k}{r} \sinh^{-1} x}-\frac{2 M}{r},\label{0.7}\\
        &&f_{Bur}(r)=e^{\frac{4 \pi  k }{r}(1+x) \arctan x}(1+x)^{-\frac{4 \pi  k }{r}(1+x)} (1+x^2)^{\frac{2 \pi  k}{r} (x-1)}-\frac{2 M}{r},\label{1.4}\\
        &&f_{Bro}(r)=e^{\frac{4 \pi  k x^2 }{h}\, _2F_1(\frac{2}{3},1;\frac{5}{3};-\frac{r^3}{h^3})}(1+x^3)^{-\frac{8 \pi  k}{3 r}}-\frac{2 M}{r},\label{2.3}\\
        &&f_{Moo}(r)=e^{\frac{16 \pi k}{\sqrt{3} h} \arctan\frac{2 \sqrt{x}-1}{\sqrt{3}}}(1+x^{3/2})^{-\frac{16 \pi k}{3r}}(\frac{1+x-\sqrt{x}}{1+x+2\sqrt{x}})^{-\frac{8 \pi k}{3h}}-\frac{2 M}{r},\label{2.8}
    \end{eqnarray}
\end{subequations}
where $M$ is the mass of the supermassive black hole, $_{2}F_{1}$ represents the hypergeometric function and $k=h^3 \rho_0 $ can be used to give an estimation of the dark matter mass.

The various matter fields in astrophysical galaxies around the supermassive black holes also influence the gravitational lensing, and a comprehensive study including dark matter halo and all the baryonic matter in galaxies is extremely complicated. Apart from dark matter halos, there are nearby luminous stars, interstellar gas, dust, and plasma mediums. In this paper, the effect of luminous matter on the deflection of light in the Milky Way and Andromeda galaxies is roughly investigated in Appendix \ref{appendixA}. The gravitational effects of other matters are also of interest and will be explored in future studies.

\section{Theoretical Treatment \label{gravitational lensing}}

\begin{figure}
	\centering
	\includegraphics[width=0.75\textwidth]{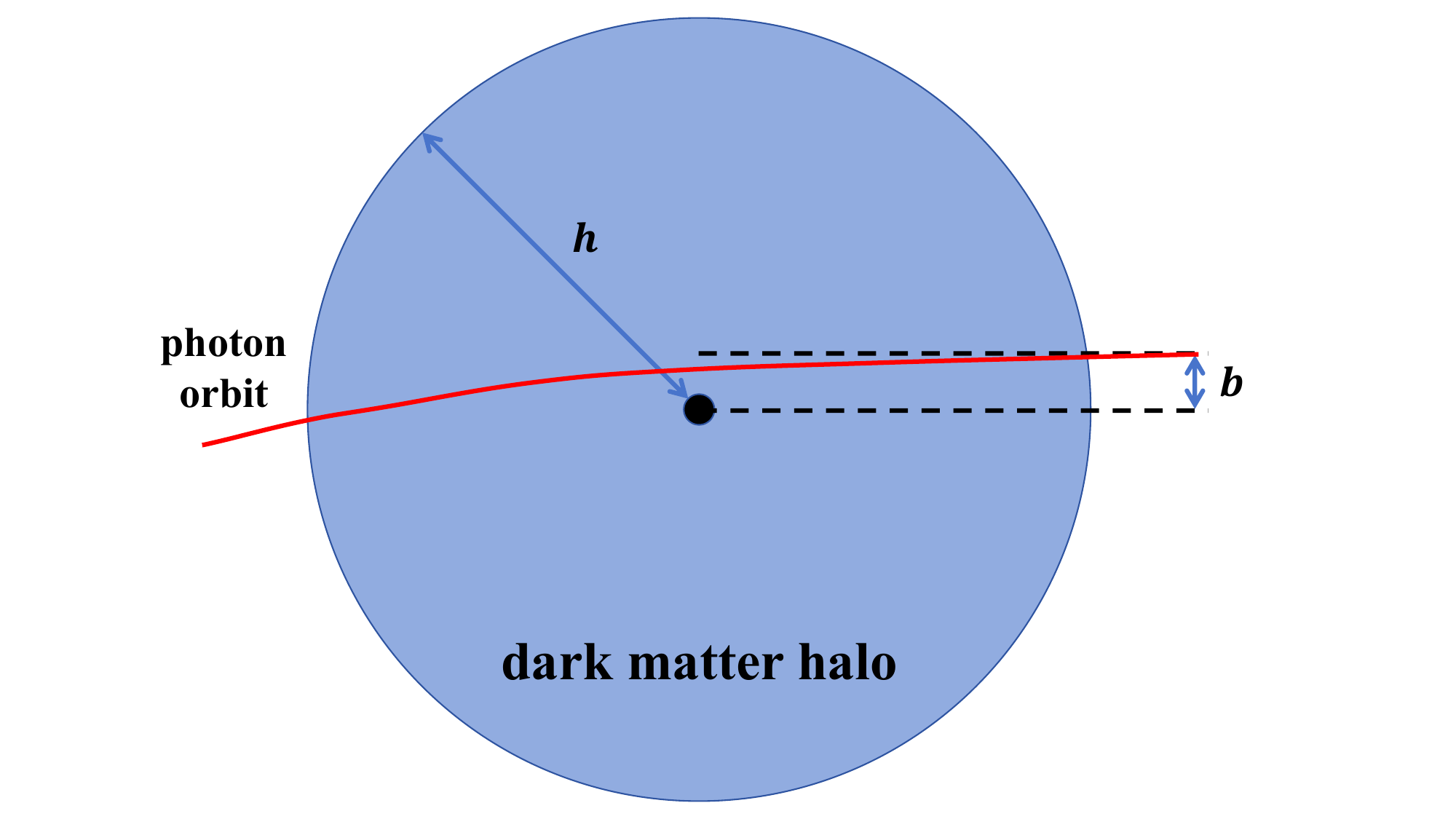}
	\caption{This figure illustrates the gravitational deflection of light in the presence of dark matter. The dark matter in the galaxy forms a halo structure, which encloses the supermassive black hole in the Galactic center. The $b$ labels the impact parameter in the gravitational lensing, and $h$ is the characteristic scale of the dark matter halo. In this work, we assume the horizon of a supermassive black hole, impact parameter, and dark matter halo scale satisfies $r_{H} \sim M \ll b \ll h$.}
    \label{figure 1}
\end{figure}

In this section, we use the generalized Gibbons and Werner (GW) method to calculate the gravitational deflection angle in asymptotically flat spacetimes. At the same time, we obtain the angular positions of lensed images and the Einstein ring angular radius by solving the famous lens equation. Notably, we assume the horizon of a supermassive black hole, impact parameter, and dark matter halo scale satisfies $r_{H} \sim M \ll b \ll h$. The gravitational deflection of light under this assumption is illustrated in Fig. \ref{figure 1}.

\subsection{Gravitational deflection angle}\label{3a}

According to Einstein's general theory of relativity, gravitational interaction is a geometric effect, so analyzing gravitational bending and gravitational lensing through new techniques of geometry and topology is possible. The Gauss-Bonnet theorem was first proposed by G. W. Gibbons and M. C. Werner to calculate the gravitational deflection angle of particles \cite{Gibbons:2008rj}. In their work, the gravitational deflection angle is calculated by a surface integral in the optical geometry
\begin{equation}\label{0.01}
    \alpha  =-\int \int _DK dS.
\end{equation}
with $K$ to be the Gaussian curvature of optical geometry.

Recently, this approach has been applied to large numbers of static and stationary gravitational systems, and several improved schemes are proposed \cite{Ishihara:2016vdc, Li:2020wvn, Werner:2012rc, Takizawa:2020egm, Huang:2022soh, Huang:2023bto}. Notably, Y. Huang and Z. Cao proposed a generalized Gibbons and Werner (GM) method that works well for asymptotically flat and nonflat spacetimes, and it deeply simplify the calculation of the deflection angle, seeing Ref. \cite{Huang:2022soh, Huang:2023bto} for details. In this paper, we will take this approach to carry out the calculation. Based on Huang's derivation, the gravitational deflection angle is reduced from Eq. (\ref{0.01}) as
\begin{equation}
    \alpha  =\int_{\phi_S }^{\phi_R } [1+H(r_{\gamma} )] \, d\phi, \label{0.3}
\end{equation}
with the function $H(r)$ defined by
\begin{align}
    H(r)=-\frac{1}{2 \sqrt{g }}\frac{\partial g^{OP} _{\phi \phi }}{\partial r}. \label{1.0}
\end{align}
The above expressions are derived by using the Gauss-Bonnet theorem in a two-dimensional manifold $M^{(g 2)}$, corresponding to the equatorial plane of optical space for static and spherically symmetric spacetimes(SSS) and $\phi_{R}$, $\phi_{S}$ in Eq. (\ref{0.3}) are the azimuthal angle for observer and light source. The two-dimensional optical space metric is usually constructed from the SSS spacetime metric by constraining $ds^{2}=0$ and $\theta =\pi /2$
\begin{equation}
    {dt}^2=g^{OP} _{rr}(r){dr}^2+g^{OP} _{\phi \phi }(r){d\phi }^2=\frac{1}{{f(r)}^2} {dr}^2 +\frac{r^2}{f(r)}{d\phi }^2,
\end{equation}
where $g$ in Eq. (\ref{1.0}) is the determinant of the above optical metric. The $r_\gamma $ denotes the radial coordinate of photon orbit from the light source to the observer, which can be determined by the orbital equation
\begin{equation}
    (\frac{du}{d\phi })^2=\frac{1}{b^2}-u^2 F(u), \label{0.8}
\end{equation}
where the variable $u$ is defined as $u = 1/r$, and $F(u)=f(1/u)$ is the metric of SSS spacetime.

Applying the famous Gauss-Bonnet theorem in the two-dimensional optical space of black holes, the Gibbons and Werner's method and its generalizations have achieved a great success in a number of gravitational systems, such as the rotational black holes and wormholes \cite{Werner:2012rc, Ovgun:2018tua, Jusufi:2018jof, Li:2021xhy, Liu:2022lfb}, the finite distance deflections \cite{Ishihara:2016vdc, Ishihara:2016sfv, Ono:2017pie, Li:2019qyb}, the spacetimes in the presence of plasma and dark matter \cite{Ovgun:2018oxk, Pantig:2022toh, Crisnejo:2018uyn, Crisnejo:2018ppm, Crisnejo:2019ril}, and asymptotically non-flat spacetimes \cite{Takizawa:2020egm, Li:2020wvn, Huang:2022soh, Takizawa:2023izb, Takizawa:2021jxa}. The results obtained from these methods are consistent with traditional approaches in the weak deflection cases \cite{Li:2019mqw, Ono:2019hkw, Kumaran:2021rgj, Sanchez:2023ckq, Li:2022cpu}.

\subsection{The lens equation and einstein ring}
\begin{figure}
	\centering
	\includegraphics[width=0.95\textwidth]{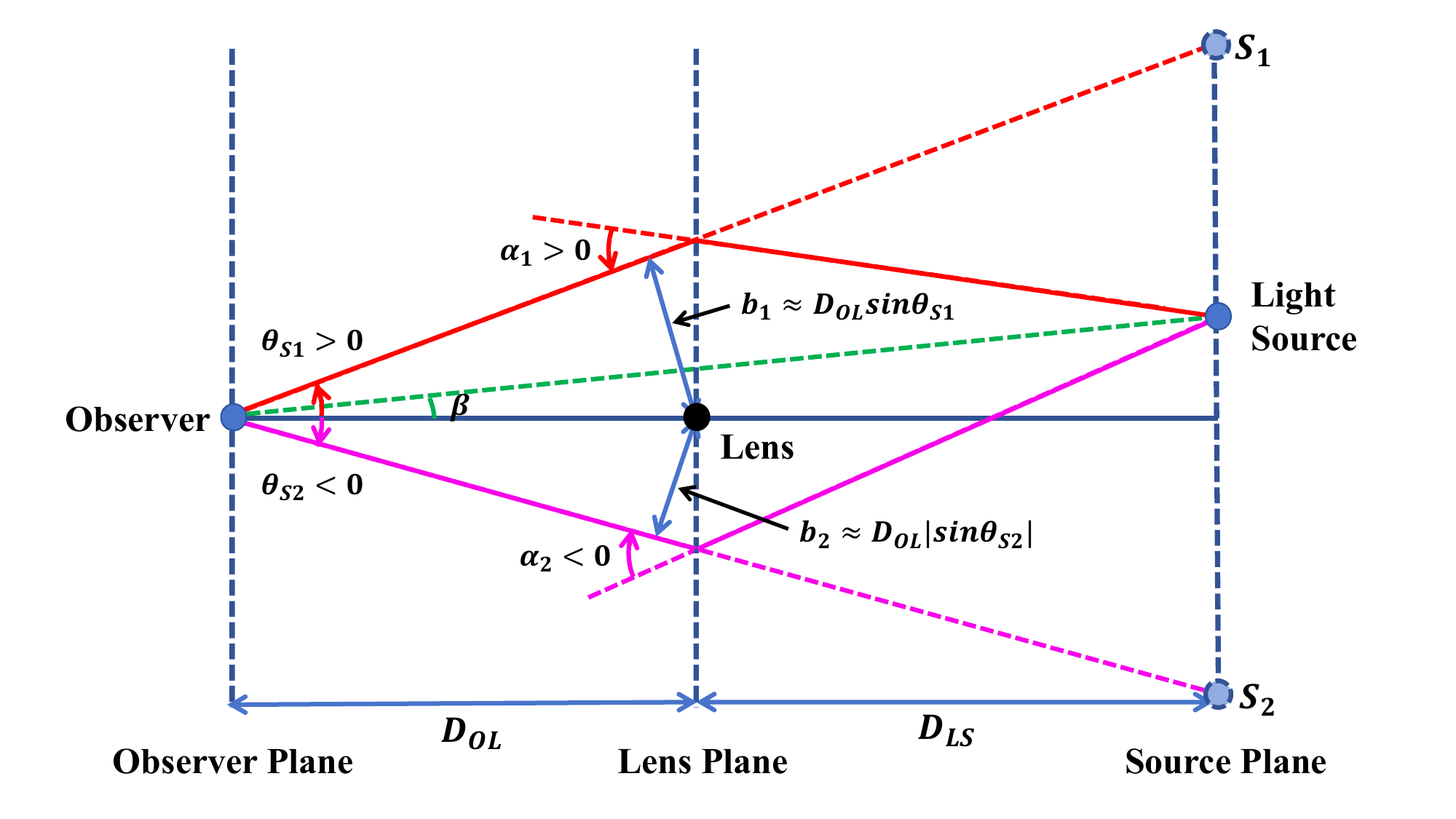}
	\caption{This figure illustrates the light propagation in the gravitational lensing. The locations of the observer, luminous light source, and the central supermassive black hole (which acts as a gravitational lens) are labeled respectively. The gravitational deflection angle of light $\alpha$, angular position of the light source $\beta$, angular positions of the lensed images $\theta_{S1}$, $\theta_{S2}$, and impact parameter $b$ have been shown in the figure. Notably, the angles $\theta_{S1}$, $\alpha_{1}$, $\beta$ in the clockwise direction are assigned to be positive, and the angles $\theta_{S2}$, $\alpha_{2}$ in the clockwise direction are assigned to be negative.}\label{figure x}
\end{figure}

In astrophysical observations, light beams emitted from remote luminous sources can be deflected and converged due to the central supermassive black hole, which plays the role of a gravitational lens. Sometimes, luminous light sources may exhibit multiple images after a gravitational lens, and the Einstein ring is such example. The angular radius of the Einstein ring is usually achieved by solving the lens equation. Many gravitational lens equations have been proposed in astrophysical studies \cite{Virbhadra:1999nm, Dabrowski:1998ac, Frittelli:1998hr, Perlick:2003vg, Bozza:2006sn}. In our work, we choose a famous lens equation is given by V. Bozza \cite{Bozza:2008ev},
\begin{equation}
    D_{OS}\cdot \text{tan$\beta $}=\frac{D_{OL} \cdot\text{sin$\theta_S$}-D_{LS}\cdot \text{sin$(\alpha -\theta_S)$}}{\text{cos$(\alpha -\theta_S)$}}.\label{0.4}
\end{equation}
Here, $D_{OS}$ is the distance between the observer and the source plane, $D_{OL}$ is the distance between the observer and the lens plane, $D_{LS}$ is the distance between the lens plane and the source plane, angle $\beta $ represents the precise angular position of the luminous source, and angle $\theta_{S}$ is the visual angular position of the lens image as seen by a distant observer. The actual position of a luminous source and its visual images produced by a gravitational lens are shown in figure \ref{figure x}. In many cases, the lensed astrophysical luminous sources could have multiple images. The figure illustrated the most general two-image cases in gravitational lensing observations. In the source plane, the image $S_{1}$ and the light source are located on the same side, while the image $S_{2}$ are located on the other side of the light source (they are separated by the central supermassive black hole). We adopt the following convention for angle $\alpha$, $\beta$, and $\theta_{S}$: the angles in the counter-clockwise direction to be positive, and the angles in the clockwise direction to be negative. Under such convention, the observables in the gravitational lensing can be applied to the same lens equation (\ref{0.4}), regardless of which direction the light beams are deflected during the propagation.

In most of the gravitational lensing, the light beams are emitted from distinct sources, which makes the gravitational deflection angle very small and the weak gravitational deflection limit available. In the weak gravitational deflection limit, the approximations $tan\beta\approx \beta $, $sin\theta_{S} \approx \theta_{S} $, and $sin (\alpha -\theta_{S} )\approx \alpha -\theta_{S} $ can be made. And the relation $D_{OS}=D_{OL}+D_{LS}$ reduces the gravitational lens Eq. (\ref{0.4}) to
\begin{equation}\label{3.4}
    \beta =\theta_S -\frac{D_{LS}}{D_{OS}}\cdot \alpha .
\end{equation}

In theory, given any value of $\beta $, we can find the corresponding $\theta _S$. However, we are of great interest to the Einstein ring angular radius. By taking $\beta =0$ to calculate the angular radius of Einstein ring, the formula is obtained
\begin{equation}
    \theta _E=\frac{D_{LS}}{D_{OS}}\cdot \alpha.\label{0.5}
\end{equation}
In addition, in the weak deflection cases, the impact parameter $b$ is satisfied
\begin{equation}
    b\approx D_{OL} \cdot \text{sin$\theta _E $}.\label{0.6}
\end{equation}
The angular radius $\theta _E$ of Einstein's ring can be solved by Eqs. (\ref{0.5}) and (\ref{0.6}) as long as the gravitational deflection angle $\alpha $ is known.

\section{Calculation of Gravitational Deflection Angles \label{gravitational deflection angle}}

In this section, we present the results and discussions on the gravitational deflection angle for black hole surrounded by a dark matter halo. Besides, we mainly focused on the weak gravitational deflection limit, so that the gravitational deflection angles are dominantly contributed by the leading-order $M/b$ and $k/b$ terms. The subsections specify the gravitational deflection angles for the dark matter distributions given by Beta, Burkert, Brownstein, and Moore models.

\subsection{Beta model}\label{Beta}
In this subsection, we calculate the gravitational deflection angle of black hole surrounded by Beta model dark matter halo. The metric of $M^{(g 2)}$ corresponding to the equatorial plane of optical space for dark matter distribution is
\begin{equation}
    \text{d\textit{l}}^2=\frac{1}{f^2(r)}{dr}^2+\frac{r^2}{f(r)}{d\phi }^2.\label{0.9}
\end{equation}
According to the generalized GW method described in subsection \ref{3a}, the gravitational deflection angle can be calculated through the integration of function $H(r)$. Substituting Eqs. (\ref{0.7}) and (\ref{0.9}) into Eq. (\ref{1.0}) leads to
\begin{equation}
    H(r)=-1+\frac{2 M}{r}+\frac{4 \pi k}{r}\biggl[2\mathrm{arcsinh}\frac{r}{h}-\frac{1}{\sqrt{1+(\frac{h}{r})^2}}\biggr] 
    +\mathcal{O}(M^2,k^2,kM).
\end{equation}
Considering the motion on the equatorial plane, the photon orbit equation can be represented by Eq. (\ref{0.8})
\begin{equation}
    (\frac{du}{d\phi })^2=\frac{1}{b^2}-u^2+2 u^3 (M-4 \pi  k \ln \frac{h u}{2})+\mathcal{O}(M^2,k^2,kM),
\end{equation}
with a new variable $u=1/r$ defined in photon orbit. Thus, we can get the perturbation solution
\begin{equation}
	\begin{split}\label{1.1} 
    u =& \frac{\sin \phi }{b}+\frac{M \left(1+\cos ^2\phi \right)}{b^2}+\frac{2 \pi  k}{b^2}\biggl[1+3 \ln \frac{2 b \csc \phi }{h}+(1+\ln \frac{2 b \csc \phi }{h})\cos 2 \phi 
    \\
    & +4 \cos \phi  \ln \tan \frac{\phi }{2}\biggr] +\mathcal{O}(M^2,k^2,k M).
    \end{split} 
\end{equation}
For both the light source and the receiver are at infinity, $u=1/r\to 0$, the azimuthal angles $\phi _{R}$ and $\phi _{S}$ can be approximated by $\phi _R\approx \pi+\alpha \approx \pi$ and $\phi _S\approx 0 $. We calculate the gravitational deflection angle through Eq. (\ref{0.3})
\begin{equation}
    \label{3.3}
    \alpha=\int_{0 }^{\pi } [1+H(r_{\gamma} )] \, d\phi
        =\frac{4 M}{b}+\frac{4 \pi  k}{h}(\pi-z^3)+\mathcal{O}(M^2,k^2,kM),
\end{equation}
where $z=b/h$ and the $r_\gamma =1/u$ have been used with Eq. (\ref{1.1}). The higher-order terms are not presented here. With the gravitational deflection angle, we can solve the lens equation to obtain the Einstein ring angular radius, which is studied in section \ref{ring}.

\subsection{Burkert model}\label{Bur}
We present calculation of the gravitational deflection angle for a black hole surrounded by the Burkert model dark matter halo in this subsection. Substitute  Eqs. (\ref{1.4}) and (\ref{0.9}) into Eq. (\ref{1.0}) to get
\begin{equation}
	\begin{split}
    H(r)=&-1+\frac{2 M}{r}+\frac{\pi k}{r}\biggl[(2-\frac{r}{h}) \ln (1+\frac{r^2}{h^2})-2 (2+\frac{r}{h}) (\arctan \frac{r}{h}-\ln (1+\frac{r}{h}))\biggr] \\
    &+\mathcal{O}(M^2,k^2,kM).
    \end{split}
\end{equation}
Since the motion is restricted on the equatorial plane, the photon orbit equation can be represented
\begin{equation}
    (\frac{du}{d\phi })^2=\frac{1}{b^2}-u^2+2 M u^3-\frac{2 \pi  k u^2 }{h}\biggl[\pi+hu(\pi -4+4\ln hu) \biggr]+\mathcal{O}(M^2,k^2,kM).
\end{equation}
The perturbation solution is
\begin{equation}
    \begin{split}\label{2.0}
    u=&\frac{\sin \phi }{b}+\frac{M (1+\cos ^2\phi )}{b^2}-\frac{\pi  k}{2 b^2 }\biggl[3 \pi -16+2 \pi\frac{b}{h}\sin\phi+12 \ln \frac{h \sin \phi }{b} 
    \\
    &+(\pi -8+4 \ln \frac{h \sin \phi }{b})\cos 2 \phi+(\pi^2 \frac{b}{h} -2\pi\phi\frac{b}{h}+16\ln \cot \frac{\phi }{2})\cos \phi\biggr]
    \\
    &+\mathcal{O}(M^2,k^2,k M).
    \end{split}
\end{equation}
When the light source and the receiver are at infinity, having employed approximations of $\phi _R\approx \pi+\alpha \approx \pi$ and $\phi _S\approx 0 $, we obtain the gravitational deflection angle through Eq. (\ref{0.3})
\begin{equation}
    \alpha=\int_{0 }^{\pi } [1+H(r_{\gamma} )] \, d\phi
        =\frac{4 M}{b}+\frac{\pi  k z^4}{18 b}(5-3 \pi+12 \ln \frac{z}{2})+\mathcal{O}(M^2,k^2,kM),
\end{equation}
where the $r_\gamma =1/u$ has been used with Eq. (\ref{2.0}).

\subsection{Brownstein model}\label{Bro}
For a black hole surrounded by the Brownstein model dark matter halo, the gravitational deflection angle can be calculated using the same scheme in the previous subsections. The function $H(r)$ can be obtained through substituting Eqs. (\ref{2.3}) and (\ref{0.9}) into Eq. (\ref{1.0})
\begin{equation}
    H(r)=-1+\frac{2 M}{r}+\frac{2\pi k}{r}\biggl[\frac{4}{3}\ln (1+\frac{r^3}{h^3})-\frac{r^3}{h^3}\,  _2F_1(\frac{2}{3},1;\frac{5}{3};-\frac{r^3}{h^3})\biggr]+\mathcal{O}(M^2,k^2,kM).
\end{equation}
The photon orbit equation on the equatorial plane can be expressed
\begin{equation}\label{0.005}
    (\frac{du}{d\phi })^2=\frac{1}{b^2}+2 M u^3-u^2 (1+\frac{16\sqrt{3} \pi ^2 k}{9 h}) (1-8 \pi  k u+8 \pi  k u \ln hu)+\mathcal{O}(M^2,k^2,kM).
\end{equation}
The perturbation solution can be gained
\begin{equation}
    \begin{split}\label{2.7}
    u=&\frac{\sin \phi }{b}+\frac{M (1+\cos ^2\phi )}{b^2}+\frac{2 \pi  k}{9 b^2}\biggl[-2 \cos \phi \big(\sqrt{3} \pi  (\pi -2 \phi )\frac{b}{h}+18\ln\cot \frac{\phi }{2}\big)
    \\
    &+18 (2+\cos 2 \phi)-9(\cos 2 \phi +3) \ln\frac{h \sin \phi }{b}-4 \sqrt{3} \pi\frac{b}{h}\sin\phi \biggr]+\mathcal{O}(M^2,k^2,k M).
    \end{split}
\end{equation}
Finally, the gravitational deflection angle is calculated from Eq. (\ref{0.3}) with the function $H(r)$ and photon orbit specified by Eqs. (\ref{0.005}) and (\ref{2.7})
\begin{equation}
    \alpha=\int_{0 }^{\pi } [1+H(r_{\gamma} )] \, d\phi
        =\frac{4 M}{b}+\frac{2 \pi  k }{9 h}(8 \sqrt{3} \pi z  -9  \text{C}_1(z))+\mathcal{O}(M^2,k^2,kM),
\end{equation}
where $r_\gamma =1/u$ has been used and $\text{C}_1(z)=z^2\int_0^{\pi } \csc ^2\phi  \, _2F_1\left(\frac{2}{3},1;\frac{5}{3};-z^3 \csc^3 \phi \right) \, \rm{d}\phi$ is a function of $z=b/h$, which could be calculated numerically. The detailed behavior of function $\text{C}_1(z)$ can be seen in Appendix \ref{appendixComparison}.

\subsection{Moore model}\label{Moo}
For the calculation of black hole surrounded by Moore model dark matter halo, through substituting Eqs. (\ref{2.8}) and (\ref{0.9}) into Eq. (\ref{1.0}), one can get
\begin{equation}
	\begin{split}
    H(r) =& -1+\frac{2 M}{r}+\frac{4\pi k}{3r}\biggl[4\ln (1+(\frac{r}{h})^{\frac{3}{2}})-\frac{2r}{h}(\sqrt{3} \arctan \frac{2 \sqrt{\frac{r}{h}}-1}{\sqrt{3}} 
    \\
    & +\mathrm{arctanh}\frac{3}{1+2\sqrt{\frac{r}{h}}+2\sqrt{\frac{h}{r}}})\biggr] +\mathcal{O}(M^2,k^2,kM).
    \end{split}
\end{equation}
On the equatorial plane, the photon orbit equation can be represented
\begin{equation}
    \begin{split}
    (\frac{du}{d\phi })^2&=\frac{1}{b^2}+u^2 \biggl[2 u (M+4 \pi  k)-\frac{8 \sqrt{3} \pi ^2 k}{3h}-1\biggr]-8 \pi  k u^3 \ln h u+\mathcal{O}(M^2,k^2,kM).
    \end{split}
\end{equation}
We can get the perturbation solution
\begin{equation}
    \begin{split}\label{3.0}
    u=&\frac{\sin \phi }{b}+\frac{M (1+\cos ^2\phi)}{b^2}-\frac{2 \pi  k}{3 b^2 }\biggl[\big(\sqrt{3}\pi (\pi -2 \phi )\frac{b}{h}+12 \ln \cot \frac{\phi }{2}\big)\cos \phi\\&-6 (2+\cos 2 \phi )+3 (3+\cos 2 \phi ) \ln \frac{h \sin \phi }{b}+2\sqrt{3}\pi\frac{b}{h}\sin \phi\biggr]+\mathcal{O}(M^2,k^2,k M).
    \end{split}
\end{equation}
The $r_\gamma =1/u$ has been used with Eq. (\ref{3.0}). Obtaining the above integration function $H(r_{\gamma})$ and photon orbit, the gravitational deflection angle can be calculated through Eq. (\ref{0.3})
\begin{equation}
    \begin{split}
    \alpha=&\int_{0 }^{\pi } [1+H(r_{\gamma} )] \, d\phi\\
        =&\frac{4 M}{b}+\frac{8\sqrt{3} \pi  k}{9 b}\biggl[2 \pi  z^{3/2}(\frac{\pi ^{3/2} \Gamma (\frac{3}{4})}{\Gamma (\frac{7}{12})^2 \Gamma (\frac{11}{12})^2 \Gamma (\frac{5}{4})}-2 \sqrt{z})-3\text{C}_2(z)+\sqrt{3}\text{C}_3(z)\biggr]\\
        &+\mathcal{O}(M^2,k^2,kM),
    \end{split}
\end{equation}
where $\text{C}_2(z)=z\int_0^{\pi } \arctan \frac{2 \sqrt{ z\csc \phi }-1}{\sqrt{3}} \, \rm{d}\phi$ and $\text{C}_3(z)=z\int_0^{\pi } \mathrm{arctanh} \frac{3 \sqrt{z \csc \phi }}{2+2 z \csc \phi +\sqrt{z \csc \phi }} \, \rm{d}\phi$ can be calculated numerically. The details of functions $\text{C}_2(z)$ and $\text{C}_3(z)$ are shown in Appendix \ref{appendixComparison}.

\subsection{Comparison of numerical results\label{cnr}}
\begin{figure}[]
	\centering
	\includegraphics[width=8cm]{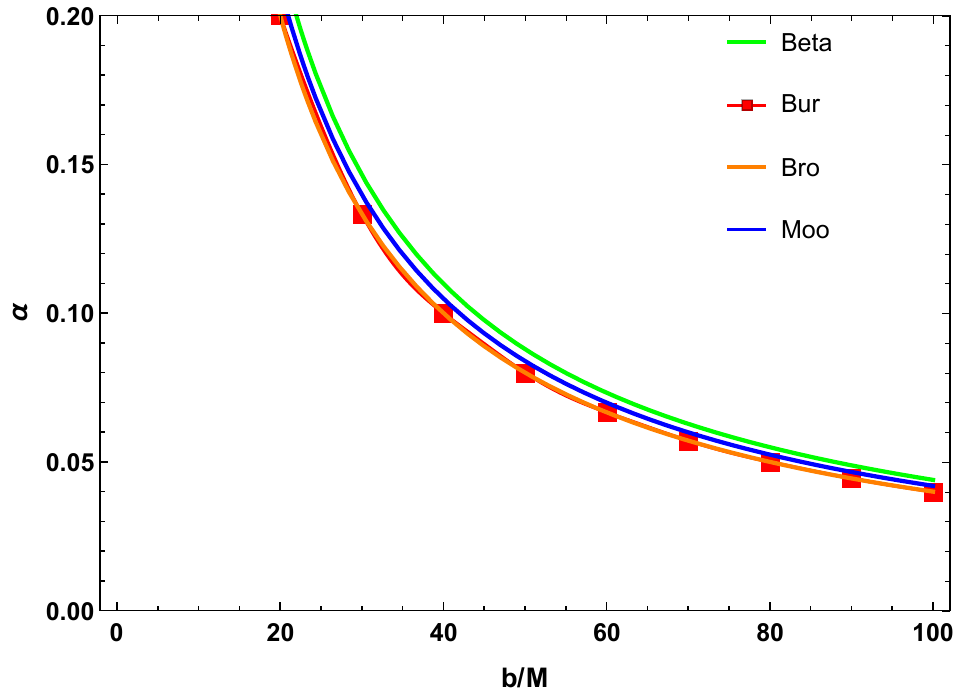}
	\caption{The gravitational deflection angles of black holes surrounded by Beta, Burkert, Brownstein, and Moore dark matter halos with different $b/M$. In this figure, we take the parameter $k=M$, and the characteristic radius of the dark matter halo is $h=100b$. The green line is the Beta model result, the red square dotted line is the Burkert model result, the orange line is the Brownstein model result, and the blue line is the Moore model result. Since the Burkert and Brownstein models results overlap, the Burkert model result is distinguished by a red square dotted line.}
	\label{figure b}
\end{figure}

\begin{figure}[]
	\centering
	\includegraphics[width=8cm]{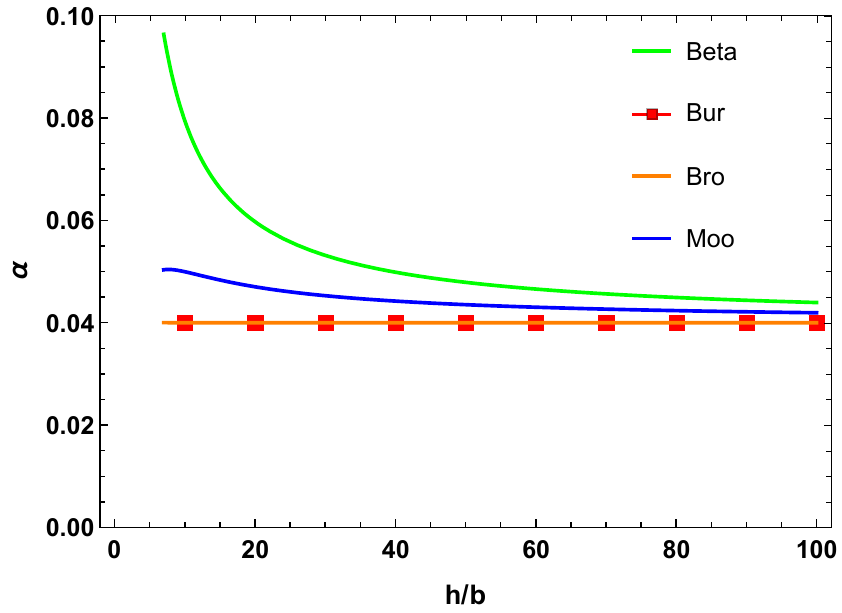}
	\caption{The gravitational deflection angles of black holes surrounded by Beta, Burkert, Brownstein, and Moore dark matter halos with different values of $h/b$. In this figure, we also take the parameter $k=M$, and the impact parameter is selected as $b=100M$. The thumbnail is an enlarged image of the Brownstein model. The green, orange, blue lines correspond to Beta, Brownstein, Moore results, and the Burkert model result is distinguished by a red square dotted line, which is the same as in Fig. \ref{figure b}}
	\label{figure h}
\end{figure}

To study the dark matter effects on gravitational deflection of black holes surrounded by dark matter halos, it would be necessary to give Comparisons between the gravitational deflection results using the four dark matter halo models. In the actual astrophysical gravitational lensing observations, the observer is usually very far from the lensed black holes. Therefore, the assumption of infinite distance observer and light source in the calculations of gravitational deflection angles displayed in subsections \ref{Beta}--\ref{Moo} is reasonable. We plot the gravitational deflection angle for infinite distance observers and light sources in Figs. \ref{figure b} and \ref{figure h}.

At different $b/M$ for both the light source and the receiver at infinity, the gravitational deflection angles of black holes surrounded by Beta, Burkert, Brownstein, and Moore dark matter halos are shown in Fig. \ref{figure b}. We take the parameter $k=M$, and the characteristic radius of the dark matter halo $h=100b$ in this figure. We can see from the figure that the gravitational deflection angle decreases with the increase of the $b/M$. In other words, when the dark matter halo parameters are fixed, with the increase of the impact parameter, the influence of the central supermassive black hole on the light decreases, resulting in the reduction of the gravitational deflection angle. Furthermore, for the same dark matter mass, dark matter halo scale, and impact parameters, the Beta dark matter halo model results in the largest gravitational deflection angle, followed by the Moore model. The deflection angle in the Burkert and Brownstein models are the smallest, and the curves of these two models are overlapped in the figure.

At different $h/b$ for both the light source and the receiver at infinity, the gravitational deflection angles of black holes surrounded by Beta, Burkert, Brownstein, and Moore dark matter halos sre shown in Fig. \ref{figure h}. We also take the parameter $k=M$, and the impact parameter is elected as $b=100M$ in this figure. With the increase of $h/b$, the gravitational deflection angles in Beta, Burkert, Brownstein, and Moore models decrease and converge to $M/b=0.04$. In other words, when the impact parameter is determined, for the Beta, Burkert, and Moore models, as the dark matter halo characteristic radius increases, the influence of the dark matter halo on gravitational deflection angle decreases. This is because, for a dark matter halo with a larger scale and constant mass, the dark matter is more sparsely distributed and the central black hole is more weakly affected by the dark matter medium. In the $h$ $\to$ $\infty$ limit, the gravitational deflection angles for different halo models converge to the $4M/b$ term (the gravitational deflection angle without the dark matter). However, the Burkert and Brownstein model curves are nearly straight lines, indicating that the dark matter halo has little effect on their gravitational deflection angles. Furthermore, for the same dark matter mass, dark matter halo scale, and impact parameters, the Beta dark matter halo model results in the largest gravitational deflection angle, followed by the Moore model. The Burkert and Brownstein models result in the smallest deflection angles, and their curves are overlapped in the figure.

\section{Angular Position Of Lensed Images and Einstein Ring}\label{ring}
In this section, we discuss some valuable observables in gravitational lensing observations. Through the gravitational lensing, the visual angular position of lensed images $\theta _S$ could be produced. Especially, the Einstein ring, a special case of gravitationally lensed images with $\beta =0$, is of more importance and deserves a detailed discussion in our work.

We mainly discuss the gravitational lensing from the supermassive black holes in our Galaxy (Sgr*A in the Milky Way), Andromeda galaxy(M31), Virgo galaxy(M87), and ESO138-G104 galaxy in the presence of dark matter halos. These black holes act as gravitational lenses and could provide typical examples in astrophysical gravitational lensing observations. Since the light sources in these gravitational lensing observations are distant from us, one can use the gravitational lens equation and the analytical gravitational deflection angle derived from the weak deflection limit to calculate the Einstein ring angular radius. The position of the lensed image is calculated numerically by using Eq. (\ref{3.4}). Further, setting $\beta = 0$, the Einstein ring angular radius of the above dark matter models can be calculated with $\theta _E=\frac{D_{LS}}{D_{OS}}\cdot \alpha$, where the gravitational deflection angle $\alpha$ of the four models can be expressed as
\begin{subequations}\label{3.5}
    \begin{eqnarray}
        \alpha _{Beta}&=&\frac{4 M}{b}+\frac{4 \pi  k}{h}(\pi-z^3),\\
        \alpha _{Bur}&=&\frac{4 M}{b}+\frac{\pi  k z^4}{18 b}(5-3 \pi+12 \ln \frac{z}{2}),\\
        \alpha _{Bro}&=&\frac{4 M}{b}+\frac{2 \pi  k }{9 h}(8 \sqrt{3} \pi z  -9  \text{C}_1(z)),\\
        \alpha _{Moo}&=&\frac{4 M}{b}+\frac{8 \pi  k}{9 b}\biggl[2 \pi  z^{3/2}(\frac{\pi ^{3/2} \Gamma (\frac{3}{4})}{\Gamma (\frac{7}{12})^2 \Gamma (\frac{11}{12})^2 \Gamma (\frac{5}{4})}-2 \sqrt{z}) -3\text{C}_2(z)+\sqrt{3}\text{C}_3(z)\biggr].
    \end{eqnarray}
\end{subequations}
Note that the Einstein ring angular radius $\theta_E$ is usually very small in the astrophysical gravitational lensing observations. The impact parameter $b$ in gravitational lensing satisfies Eq. (\ref{0.6}). Finally, the Einstein ring angular radius of a lensed luminous object can be calculated by solving Eqs. (\ref{3.5}) and (\ref{0.6}).

\begin{figure}[]
	\begin{center}
		\subfigure[$\ \beta =0$ arcsec]{\includegraphics[width=7cm]{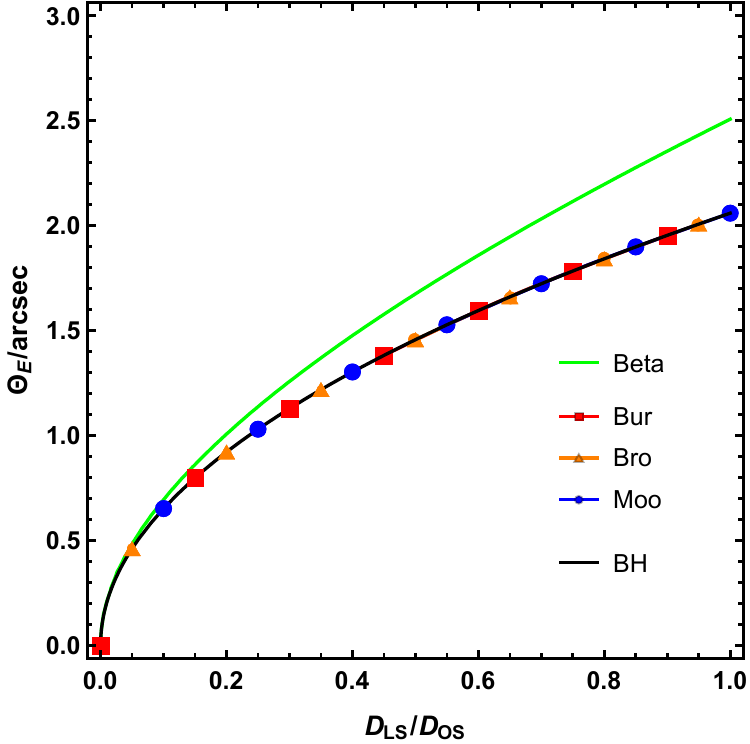}\label{figure ring M}}
		\subfigure[$\ \beta =1$ arcsec]{\includegraphics[width=7cm]{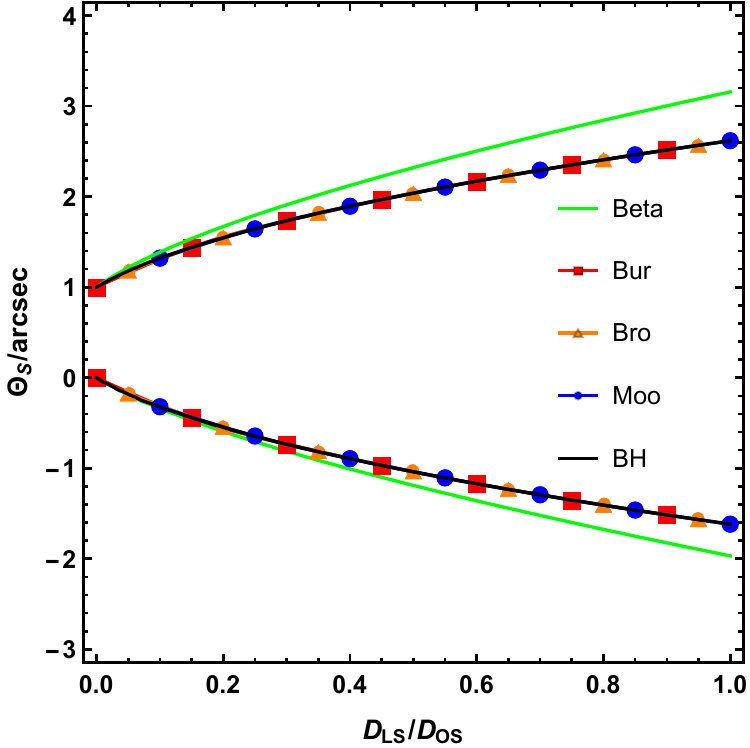}\label{figure ring1 M}}
	\end{center}
	\caption{The angular radius of the Einstein ring at $\beta =0$ (\textbf{LEFT}) and the angular positions of lensed images at $\beta =1$ arcsec (\textbf{RIGHT}) for the supermassive black hole Sgr*A in the center of the Milky Way surrounded by a dark matter halo (described by the Beta, Burkert, Brownstein, and Moore halo models). The horizontal axis labels the ratio of $D_{LS}/D_{OS}$, and the vertical axis labels the Einstein ring angular radius $\theta_E$ and the angular positions of lensed images $\theta_S$ in units of arc-second. The curve label by "BH" represents the gravitational lensing of a central black hole without dark matter. Here, the mass of the supermassive black hole is $M=4.3\times 10^6M_\odot $, the characteristic radius of the dark matter halo is $h=10.94$ kpc, and the distance between the observer and the lens plane is $D_{OL}=8.3$ kpc. The density $\rho_0$ of dark matter halos in the Milky Way are presented in Table. \ref{Table M}}
	\label{figure ring Mz}
\end{figure}

To present the numerical calculations of observables in the gravitational lensing of black holes in dark matter halos, we first consider the supermassive black hole Sgr*A in the Milky Way Galaxy. According to the recent works in the Milky Way \cite{Kafle:2014xfa, Lin:2019yux, Junior:2023xgl}, the mass of the supermassive black hole Sgr*A is $M=4.3\times 10^6M_\odot $, the distance between the observer and the lens plane is $D_{OL}=8.3$kpc, the characteristic radius of the dark matter halo is about $h=10.94$kpc, and the dark matter halos densities corresponding to models Beta, Burkert, Brownstein, and Moore are shown in Table \ref{Table M}. The angular radius of the Einstein ring and the angular positions of lensed images for this situation are drown in Fig. \ref{figure ring Mz}. The actual angular positions of the light source are chosen to be $\beta =0$ and $\beta =1$ arcsec to show the most general images configurations in the astrophysical gravitational lensing observations. The Einstein ring image is generated at $\beta =0$. At $\beta =1$ arcsec, it can produce the double images $S_1$ and $S_2$, as illustrated in Fig \ref{figure x}. The Fig. \ref{figure ring M} shows, as the $D_{LS}/D_{OS}$ increases, the Einstein ring angular radius of the Beta model increases the fastest, while the Burkert, Brownstein, and Moore models increase slower. The results show that under the same conditions, the Einstein ring angular radius of the Beta model is the largest, and the Burkert, Brownstein, and Moore models are smaller. At $\beta =1$ arcsec, as the Fig. \ref{figure ring1 M} shows, with the $D_{LS}/D_{OS}$ increases, the angular positions of lensed images have the similar tendency with the $\beta =0$ case. Note that for $\beta =0$ and $\beta =1$ arcsec, the results of Burkert, Brownstein, and Moore models almost coincide with the case of a pure black hole without a dark matter halo. It shows that the dark matter halos in Burkert, Brownstein, and Moore models contribute little to the visual angular positions of lensed images for the supermassive black hole Sgr*A in the center of the Milky Way. Furthermore, we compare the effects of luminous matter and dark matter in the Milky Way in Appendix \ref{appendixA}, and the results show that the luminous matter has a nonnegligible contribution to gravitational deflection angle compared with dark matter. In particular, Appendix \ref{appendixB} shows that the local random fluctuations of gravitational field in the Milky Way Galaxy affect the gravitational lensing variables in the order of 10 microarcseconds ($\mu$as), which can be neglected in our study.

\begin{table}[]
	\setlength{\tabcolsep}{3mm}
	\begin{center}
		\begin{tabular}[b]{c|cccc}
			\midrule[1.5pt]
			Model & Beta & Burkert & Brownstein & Moore \\
			\toprule[1pt]
			$\rho_0(10^{-3}M_\odot /pc^3)$ & 8.80 & 7.54 & 10.06 & 7.79  \\
			\midrule[1.5pt]
		\end{tabular}
	\end{center}
	\centering
	\caption{The dark matter halo density $\rho_0$ for the Beta, Burkert, Brownstein, and Moore models in the Milky Way. The numerical values listed here are selected according to the fitted values in \cite{Sofue:2020rnl, Lin:2019yux}.}
	\label{Table M}
\end{table}

\begin{figure}[]
	\begin{center}
		\subfigure[$\ \beta =0$ arcsec]{\includegraphics[width=7cm]{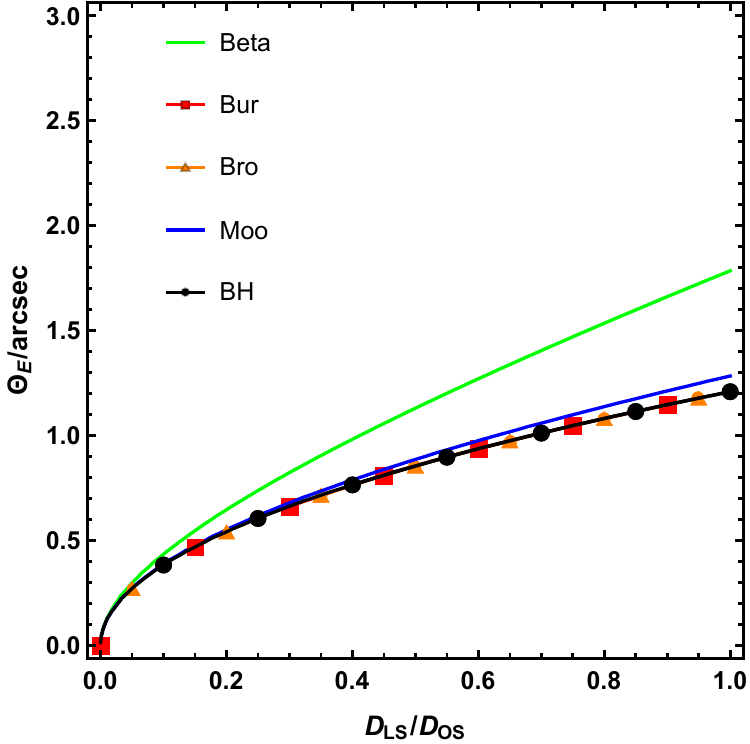}\label{figure ring M31}}
		\subfigure[$\ \beta =1$ arcsec]{\includegraphics[width=7cm]{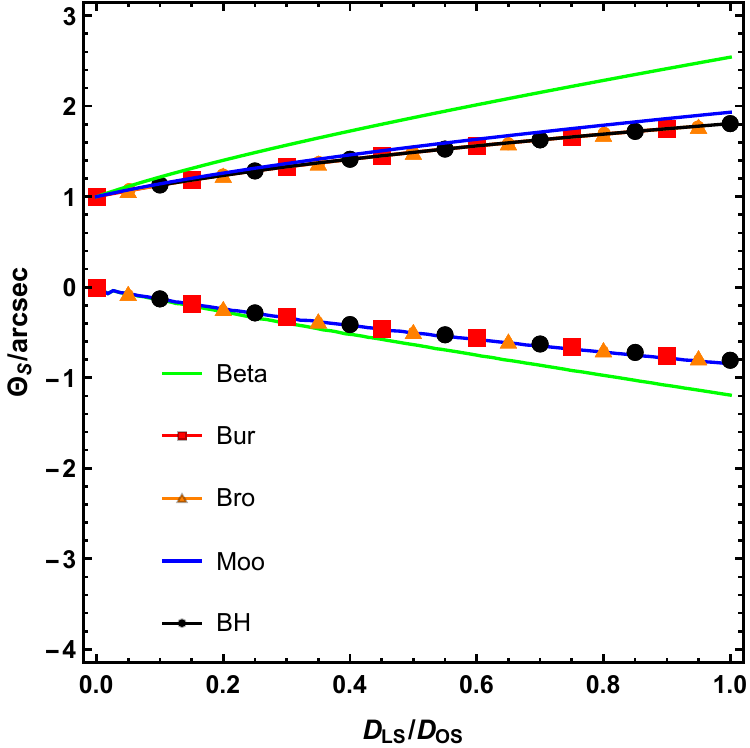}\label{figure ring1 M31}}
	\end{center}
	\caption{The angular radius of the Einstein ring at $\beta =0$ (\textbf{LEFT}) and the angular positions of lensed images at $\beta =1$ arcsec (\textbf{RIGHT}) for the supermassive black hole in the center of the Andromeda galaxy (M31) surrounded by a dark matter halo (described by the Beta, Burkert, Brownstein, and Moore halo models). The horizontal axis labels the ratio of $D_{LS}/D_{OS}$, and the vertical axis labels the Einstein ring angular radius $\theta_E$ and the angular positions of lensed images $\theta_S$ in units of arc-second. The curve label by "BH" represents the gravitational lensing of a central black hole without dark matter. Here, the mass of the supermassive black hole is $M=1.4\times 10^8M_\odot $, and the distance between the observer and the lens plane is $D_{OL}=785$ kpc. The characteristic radius $h$ and density $\rho_0$ of dark matter halos in Andromeda galaxy (M31) are presented in Table \ref{Table M31}.}
	\label{figure ring M31z}
\end{figure}
\begin{table}[]
	\setlength{\tabcolsep}{3mm}
	\begin{center}
		\begin{tabular}[b]{c|cccc}
			\midrule[1.5pt]
			Model & Beta & Burkert & Brownstein & Moore \\
			\toprule[1pt]
			$\rho_0(10^{-3}M_\odot /pc^3)$ & 3.10 & 5.37 & 1.82 & 3.01  \\
			$h(kpc)$ & 26.23 & 22.93 & 30.12 & 29.57  \\
			\midrule[1.5pt]
		\end{tabular}
	\end{center}
	\centering
	\caption{The dark matter halo characteristic radius $h$ and density $\rho_0$ for the Beta, Burkert, Brownstein, and Moore models in the Andromeda galaxy (M31). The numerical values listed here are selected according to the fitted values in \cite{Boshkayev:2022vpn}.}
	\label{Table M31}
\end{table}

Besides the Sgr*A in the Milky Way Galaxy, it is also important to see the gravitational lensing of supermassive black hole in Andromeda galaxy (M31), which could provide an interesting example. According to the recent works in the Andromeda galaxy (M31) \cite{Bender:2005rq, Boshkayev:2022vpn, Ng:2019gch}, the mass of the supermassive black hole is $M=1.4\times 10^8M_\odot $, the distance between the observer and the lens plane is $D_{OL}=785$ kpc, which is exactly the distance between the earth and Andromeda galaxy (M31). The characteristic radius $h$ and density $\rho_0$ of dark matter halos corresponding to the Beta, Burkert, Brownstein, and Moore models are given in Table \ref{Table M31}. Based on the gravitational deflection angle in Eq. (\ref{3.5}), the angular positions of lensed images for the supermassive black hole surrounded by a dark matter halo in the center of the Andromeda galaxy (M31) are shown in Fig. \ref{figure ring M31z}. The actual angular positions of the light source are chosen to be $\beta =0$ and $\beta =1$ arcsec to show the most general images configurations in the astrophysical gravitational lensing observations. As the Fig. \ref{figure ring M31} shows, with the $D_{LS}/D_{OS}$ increases, the Einstein ring angular radius of the Beta model increases the fastest, the Moore model result model the second, while the result in Burkert and Brownstein models the slowest. The results show that under the same conditions, the Einstein ring angular radius of the Beta model is the largest. At $\beta =1$ arcsec, as the Fig. \ref{figure ring1 M31} shows, with the $D_{LS}/D_{OS}$ increases, the angular positions of lensed images have a similar tendency with the $\beta =0$ case. Note that for $\beta =0$ and $\beta =1$ arcsec, the results of Burkert and Brownstein models almost coincide with the case of a black hole without dark matter. It shows that the dark matter halos in Burkert and Brownstein models contribute little to the visual angular positions of lensed images for the supermassive black hole in the center of the Andromeda galaxy (M31). As can be seen from Figs. \ref{figure ring Mz} and \ref{figure ring M31z}, to see a larger image of the Einstein ring, the distance between the lens plane and the source plane $D_{LS}$ should be as large as possible. In addition, we also compare the effects of luminous matter and dark matter in the Andromeda galaxy in Appendix \ref{appendixA}. Unlike the case of the Milky Way galaxy, the dark matter halo has greater influences than luminous matter in the gravitational deflection angle. 

Furthermore, it is also worthy to see the magnitude of Einstein rings in the gravitational lensing of supermassive black holes in different galaxies in the presence of dark matter halos. Here, we restrict ourselves to the Burkert halo model to show the Einstein ring angular radius for gravitational lensing of the Milky Way Galaxy, Andromeda galaxy(M31), Virgo galaxy(M87), and ESO138-G014 galaxy. In the Virgo galaxy(M87), the mass of the supermassive black hole is $M=6.5\times 10^9M_\odot $, the distance between the observer and the lens plane is $D_{OL}=16.8$Mpc, the characteristic radius and density of the dark matter halo are $h=91.2$kpc, $\rho _0=6.9\times 10^6M_\odot/kpc^3$ \cite{Jusufi:2019nrn, EventHorizonTelescope:2021bee, Ru-Sen Lu:2023}. In the ESO138-G104 galaxy, the total mass of Hydrogen Intensity in the 21kpc range is $M=4.6\times 10^9M_\odot $, the distance between the observer and the lens plane is $D_{OL}=18.57$Mpc, the characteristic radius and density of the dark matter halo are $h=7.5$kpc, $\rho _0=1.3\times 10^7M_\odot/\text{kpc}^3$ \cite{Hashim:2014mka, Hashim:2013}. As shown in Table \ref{Table galaxy}, by taking $D_{LS}/D_{OS}$ as $0.2$, $0.5$, and $0.8$ respectively, we can obtain the angular radius values of Einstein rings in gravitational lensing of these galaxies. And we can observe that the gravitational lensing of Milky Way results in the largest angular radius of the Einstein ring, followed by those for Virgo galaxy(M87), the ESO138-G104 galaxy's results are smaller, and the Einstein rings for gravitational lensing of Andromeda galaxy(M31) are smallest. In other words, we can observe the largest Einstein ring images from the gravitational lensing observations of Milky Way Galaxy (the Sgr*A gravitational lensing observations).

\begin{table}[]
	\setlength{\tabcolsep}{2mm}
	\begin{center}
		\begin{tabular}[b]{c|cccc}
			\midrule[1.5pt]
			$D_{LS}/D_{OS}$ & the Milky Way & Andromeda galaxy(M31) & Virgo galaxy(M87) & ESO138-G104 \\
			\toprule[1pt]
			0.2 & 0.92 & 0.54 & 0.79 & 0.64  \\
			0.5 & 1.45 & 0.85 & 1.26 & 1.01  \\
			0.8 & 1.84 & 1.08 & 1.59 & 1.27 \\
			\midrule[1.5pt]
		\end{tabular}
	\end{center}
	\centering
	\caption{Calculated data of the Einstein ring angular radius within the Burkert model in the gravitational lensing of Milky Way, Andromeda galaxy(M31), Virgo galaxy(M87), and ESO138-G014 galaxy. We have taken $D_{LS}/D_{OS}$ as $0.2$, $0.5$, and $0.8$ respectively.}
	\label{Table galaxy}
\end{table}

\section{Conclusion and Perspectives\label{concl}}

In this work, we study the gravitational lensing of central black holes surrounded by dark matter halos. Instead of treating the dark matter effects within simplified density models, we choose several widely adopted phenomenological dark matter halo models (which are the Beta, Burkert, Brownstein, and Moore models) in the current study. These phenomenological dark matter models can give precise dark matter distributions in astrophysical galaxies and galaxy clusters. The gravitational deflection angles of light for black holes in such dark matter halos are derived analytically using a generalized Gibbons-Werner approach, and some important observables (including the visual angular positions of lensed images and the Einstein ring) in gravitational lensing observations are calculated numerically using the gravitational lens equation.  Particularly, in the calculations of visual angular positions and Einstein rings, the supermassive black holes in our Galaxy (the Sgr*A in Milky Way), Andromeda galaxy (M31), Virgo galaxy (M87), and ESO138-G014 galaxy have been chosen as typical examples, with the fitted values of dark matter halos taken into numerical calculations. 

The analytical results show that the dark matter halos described by four models all contribute to the gravitational deflection angle as leading order effects, as seen by our expansion. The dark matter contributions to the gravitational deflection angle are proportional to the dark matter halo characteristic mass $k=h^3 \rho _0$, and they decrease for larger dark matter halo characteristic radius. In the $h\rightarrow \infty $ limit, the gravitational deflection angles for different halo models all converge to the gravitational deflection angle without dark matter. The numerical results show that,  the Beta dark matter model has non-negligible influences in gravitational deflection angle, the visual angular position of images and the Einstein ring. Under the same conditions, the visual angular position and Einstein ring angular radius of the Beta model are the largest, followed by the Moore model, and results in the Burkert and Brownstein models are smaller. The dark matter halos described by the Burkert and Brownstein models only have very small influences on the angular position of images and Einstein ring angular radius. To compare the dark matter effects and luminous matter effects on gravitational lensing, a simple analysis is also carried out in the Appendix based on the gravitational potential of various matter fields, which shows that the luminous matter in the Milky Way has greater influences than the Andromeda galaxy M31. A more detailed comparison of these effects deserves a subsequent study.

We have concentrated on one of the important effects on galactic supermassive black holes in gravitational lensing observations --- the dark matter halo effects. The interactions coming from other matter fields are omitted for simplicity. However, the various matter fields in astrophysical galaxies around the supermassive black holes are extremely complicated. Apart from dark matter halos, there are nearby luminous stars, as well as interstellar gas, dust, and plasma mediums. They are all interplayed with the central supermassive black holes in the gravitational lensing observations, which also deserve extensive studies. In principle, our approach (which utilizes the effective spacetime metric produced by the central black hole and the surrounding matter fields and employs the generalized Gibbons and Werner approach on gravitational deflection angles) could effectively treat these effects in the gravitational lensing \footnote{The gravitational lensing of black holes in plasma medium also has stimulated a number of works in recent years \cite{Chowdhuri:2020ipb,Li:2023esz, Molla:2022izk, Javed:2021ymu, Crisnejo:2018ppm, Crisnejo:2018uyn}.}, as long as the matter distributions have a spherical symmetry. We hope that studies in this direction could inspire more interesting works in the near future.

\appendix 
\section{Comparing the galactic dark matter and luminous matter effects on deflection angles} \label{appendixA}

In addition to dark matter, there are a number of luminous matter in galaxies that also contribute to the gravitational deflection of photons. This raises an intriguing question, which has a greater influence on the gravitational deflection of photons, the dark matter or luminous matter? To give a qualitative understanding of the influences from dark matter and luminous matter, we begin with the analysis of the deflection angle for photon orbits through gravitational potential $\Phi$ \footnote{The distributions of luminous matter in galaxies are not always spherically symmetric (such as the disk distribution), which obstructs the application of Gauss-Bonnet theorem in spherically symmetric spacetime introduced in section \ref{3a}. Instead, we conduct an analysis on gravitational deflection angles using the gravitational potential.}. The deflection angle is closely connected to the gravitational potential via
\begin{equation}
     \alpha =2\int \nabla _{\bot }\Phi  \, ds.
\end{equation}
where $ \nabla _{\bot }\Phi$ is the gradient of gravitational potential in the transverse direction (the direction transverse to the photon propagation). By calculating the gravitational potential produced by dark matter and luminous matter, their contributions to gravitational deflection angles can be compared indirectly. Using Poisson's equation, the gravitational potentials of the Beta, Burkert, Brownstein, and Moore dark matter models are expressed as
\begin{subequations}
    \begin{eqnarray}
        &&\Phi  _{Beta}(r)=\frac{4 \pi k G}{h}(\mathrm{arctanh}\sqrt{1+x^2}-\mathrm{arccoth}\frac{1}{\sqrt{1+x^2}}+\frac{1}{x}\mathrm{arcsinh}x),\\
        &&\Phi  _{Bur}(r)=\frac{\pi  k G}{r}[(x-1) \ln (1+x^2)+2 (x+1) (\mathrm{arctan}x-\ln (x+1))],\\ 
        &&\Phi  _{Bro}(r)=\frac{2 \pi  k G}{3 r}[3 x^3 \, _2F_1(\frac{2}{3},1;\frac{5}{3};-x^3)-2 \ln (1+x^3)],\\
        &&\Phi  _{Moo}(r)=\frac{4 \pi  k G}{3 r}[2 \ln (1+x^{3/2})-2x (\ln (1+\sqrt{x})+\sqrt{3} \mathrm{arctan}\frac{2 \sqrt{x}-1}{\sqrt{3}})].
    \end{eqnarray}
\end{subequations}

In this appendix, we give a comparison about the dark matter and luminous matter effects on gravitational deflections in Milky Way and Andromeda galaxies. For the luminous matter around the Milky Way and Andromeda galaxies, the bulge and disk make up a large proportion of the total luminous matter mass. It is reasonable that we only consider the gravitational potential of these two parts and neglect other luminous parts. Noticeably, the potentials produced by the bulge part can also be assumed as spherically symmetric. However, in most spiral galaxies, the luminous stars are usually accumulated in the galaxy disk, which shows that the disk part cannot be treated in the spherical symmetrical formulation as the bulge part and dark matter halo. These luminous stars result in a larger gravitational potential $|\Phi|$ in the disk plane. For simplicity, we restrict ourselves to the potential produced by luminous matter (the bulge and disk parts) in the disk plane in this appendix. This enables us to give an upper estimation on the influences of galactic luminous matter on gravitational deflection angle, because the mass of luminous matter and their gravitational potential outside the disk plane could be much smaller compared with those in the disk plane (even if at the same distance $r$ to the central black hole). From Ref. \cite{Strigari:2012acq}, the gravitational potentials of the bulge and disk of the Milky Way are modeled as
\begin{subequations}
    \begin{eqnarray}
        &&\Phi _b(r)=\frac{G M_b}{r+c_0},\\
        &&\Phi _d(r)=\frac{G M_d}{r}(1-e^{-r/b_d}),
    \end{eqnarray}
\end{subequations}
where $M_b\simeq 10^{10} M_\odot$ is the total mass of the bulge, $c_0\simeq 0.6 \rm{kpc}$ is its scale length, $M_d\simeq 5\times 10^{11} M_\odot$ is the mass of the disk, and  $b_d\simeq 3 \rm{kpc}$ is the disk scale length. For Andromeda galaxy, the bulge has the same gravitational potential model as the Milky Way, and the disk's gravitational potential is modeled as \cite{Geehan:2005aq}
\begin{equation}
        \Phi ' _d(r)=\frac{2 \pi G \Sigma_0 R_d^2}{r}(1-e^{-r/R_d}),
\end{equation}
where $M_b=3.3\times 10^{10} M_\odot$ is the total mass of the bulge, $c_0= 0.61 \rm{kpc}$ is its scale radius, $\Sigma_0= 4.6\times 10^{8} M_\odot/\rm{kpc}^2$ is the disk central surface density, and  $R_d=5.4 \rm{kpc}$ is the disk scale radius. In this work, the natural unit $G=1$ has been set.

\begin{figure}[]
	\centering
	\includegraphics[width=8cm]{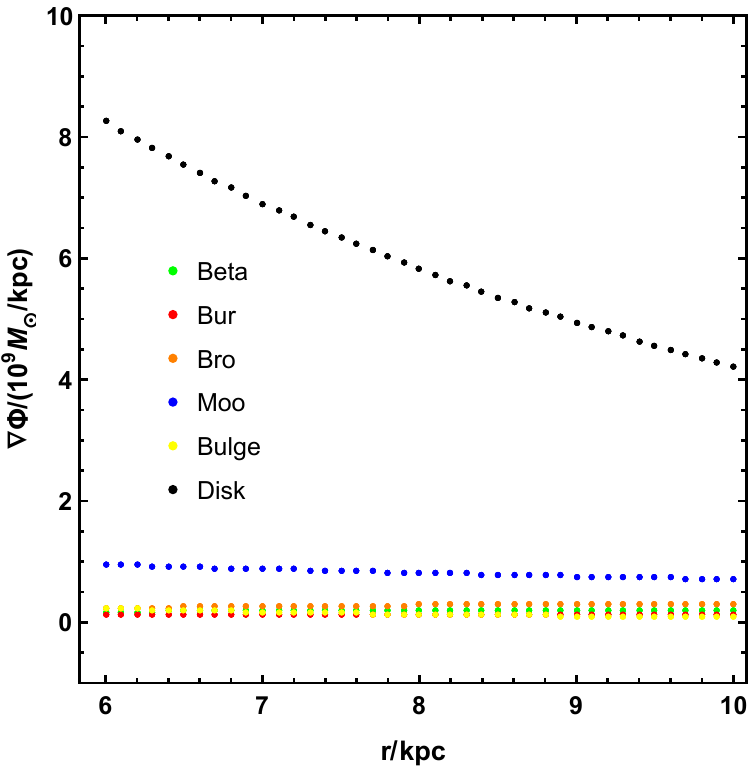}
	\caption{The gradient of gravitational potential $|\nabla \Phi|$ produced by dark matter and luminous matter in the Milky Way. This figure present the results for Beta, Burkert, Brownstein, Moore dark matter halo models, as well as the bulge and disk distributions for luminous matter. The horizontal axis is distance $r$ in unit of $\rm{kpc}$, and the vertical axis is the gravitational potential gradient in unit of $10^9 M_\odot/\rm{kpc}$.}
	\label{figure_MKP}
\end{figure}

\begin{figure}[]
	\centering
	\includegraphics[width=8cm]{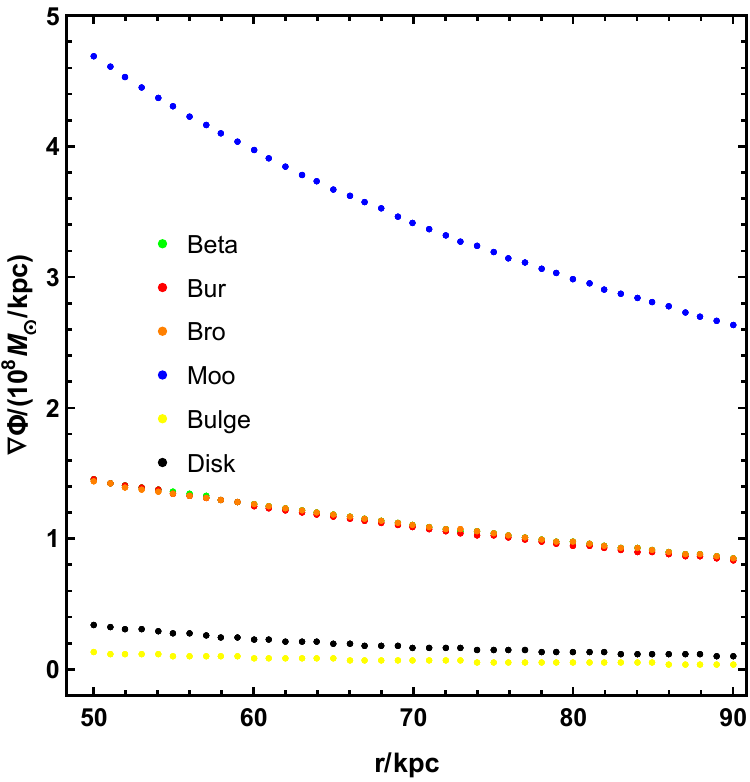}
	\caption{The gradient of gravitational potential $|\nabla \Phi|$ produced by dark matter and luminous matter in the Andromeda galaxy. This figure present the results for Beta, Burkert, Brownstein, Moore dark matter halo models, the bulge and disk distributions for luminous matter. The horizontal axis labels $r$ in unit of $\rm{kpc}$, and the vertical axis is the gravitational potential gradient in unit of $10^8 M_\odot/\rm{kpc}$.}
	\label{figure_M31P}
\end{figure}

To study the influence form bulge and disk on gravitational deflection angle, we calculate and observe how the gradient of gravitational potential changes with respect to $r$. Fig. \ref{figure_MKP} and Fig. \ref{figure_M31P} show the results for the Milky Way Galaxy and Andromeda galaxy, respectively. The numerical results of $|\nabla \Phi|$ produced by Beta, Burkert, Brownstein, Moore dark matter halo models, as well as the bulge and disk distributions for luminous matter are given in these figures. The region of distance is chosen according to the scope of galaxies as well as the distance $D_{OL}$, $D_{LS}$ in the gravitational lensing. In the Milky Way Galaxy, the distance $r$ is chosen no more than 10kpc (although the dark matter distribution scale in the Milky Way is much larger, the $D_{OL}$$\thicksim$ $D_{LS}$ $\thicksim$ 10kpc restrict the $r$ in the gravitational lensing of Milky Way). For the Andromeda Galaxy M31, the dark matter distribution could exist up to 100kpc, namely the entire dark matter halo contributes to the gravitational deflection (because $D_{OL}$ $\thicksim$ 785kpc $\gg$100kpc). Therefore, we should pick a larger distance range of $r$ for the Andromeda Galaxy. From Fig. \ref{figure_MKP}, the results show that the Milky Way disk's gravitational potential gradient is the largest, while the Burkert dark matter model's gradient is the smallest. In this case, it can be roughly inferred that the luminous matter in the Milky Way has a greater influence on the gravitational deflection of photons than the dark matter. On the other hand, in the Andromeda galaxy, the results in Fig. \ref{figure_M31P} show that the Moore model has the largest gravitational potential gradient, the bulge has the smallest, and the sum of the gravitational potential gradient of luminous matter (bulge plus disk) is smaller than that of dark matter described by four models. It is safe to concluded that luminous matter (bulge plus disk) has less effect on the deflection of photons than dark matter in Andromeda galaxy. In summary, combining the results in Fig. \ref{figure_MKP} and \ref{figure_M31P}, when studying the gravitational deflection of photons near central supermassive black holes, it is necessary to consider the effect of luminous matter in the Milky Way galaxy, while the effect of luminous matter in the Andromeda galaxy (if very high precision is not required) can be neglected.

However, it should be emphasized that the above comparisons only serve as a rough estimation of the luminous matter contributions to gravitational deflection. There are two main reasons. Firstly, our calculation for the disk part of luminous matter is restricted in the disk plane, while the luminous matter distribution and their gravitational potential outside the disk plane are less than those in the disk plane (which means we may overestimate the effects of luminous matter in the above comparisons). Secondly, for simplicity, we give numerical calculations on the total gradient of gravitational potential, rather than the transverse gradient which is more closely related to the deflection angle. Therefore, this appendix gives rough quantitative conclusions rather than high-precision predictions. A more detailed and comprehensive analysis of the luminous matter effects on gravitational lensing of supermassive black hole in the galaxy center deserve an additional study. 

\section{Effect of random fluctuations of gravitational field on gravitational lensing observable} \label{appendixB}

In an astrophysical galaxy, due to the complex (and stochastic) motions of stars and other ultra-compact objects, the gravitational field in the galaxy is not perfectly static. The local fluctuations of the gravitational field may have influences on gravitational lensing observations. This appendix provides a concise discussion of whether they have non-negligible contributions to the lensing observables calculated in our present work. 

From a pioneering work given by S. Chandrasekhar \cite{Chandrasekhar:1943ws}, the local variations of the gravitational field in time caused by the galactic objects (or the fluctuations of the galactic matter density) can be treated as a stochastic process. Inspired by this work, many studies investigate their effects on gravitational lensing \cite{Zhdanov1995, Larchenkova:2014oga, Larchenkova2017, Larchenkova2020, Yano:2012ga}. Recent studies suggested that local fluctuations of gravitational field in galaxies could lead to a “jitter” effect for apparent positions of light sources in gravitational lensing observations \cite{Larchenkova2017, Larchenkova2020}. T. I. Larchenkova et al. found that the galactic gravitational random fluctuation affects the apparent position of the light source outside the Milky River. The standard deviation of the deflection angle can reach tens of microarcseconds ($\mu as$) in the galactic center direction, while it drops to 4-6 $\mu as$ in the high silver latitude direction \cite{Larchenkova2017}. They also found that the jitter effect of apparent celestial positions of distant sources due to local fluctuations of the galaxy gravitational field can be detected when the accuracy of differential astrometric observations is around 10 $\mu as$ \cite{Larchenkova2020}. Other studies \cite{Yano:2012ga} also concluded that the influence coming from the random fluctuations of the gravitational field is at the order of microarcseconds ($\mu as$), which is not negligible in the study of microlensing \cite{Dominik:2005bp, Evans:2002uk, Dominik:1998tn}. However, the gravitational lensing observables calculated in our work (the angular radius of lensed images and Einstein ring in Sec. \ref{ring}) are at the order of arcseconds ($as$). One can safely infer that the effects of the stochastic influence of stars / local fluctuations of gravitational field on our gravitational lensing results are small. 

\section{Numerical calculations on $\text{C}_1(z)$,  $\text{C}_2(z)$ and  $\text{C}_3(z)$}\label{appendixComparison}

\begin{figure}[b]
	\begin{center}
		\subfigure[$\ \text{C}_1(z)$]{\includegraphics[width=4.8cm]{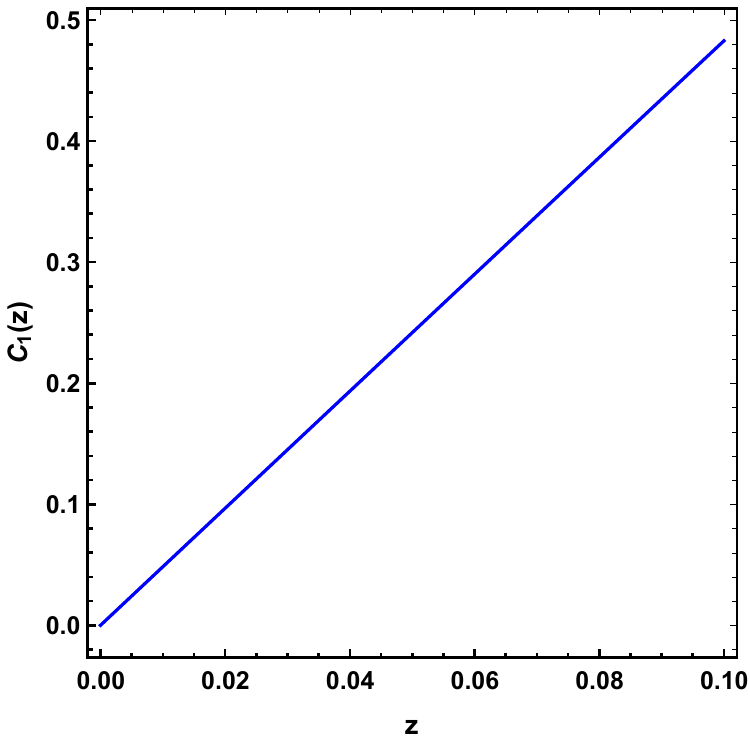}\label{figure C1}}
		\subfigure[$\ \text{C}_2(z)$]{\includegraphics[width=5.05cm]{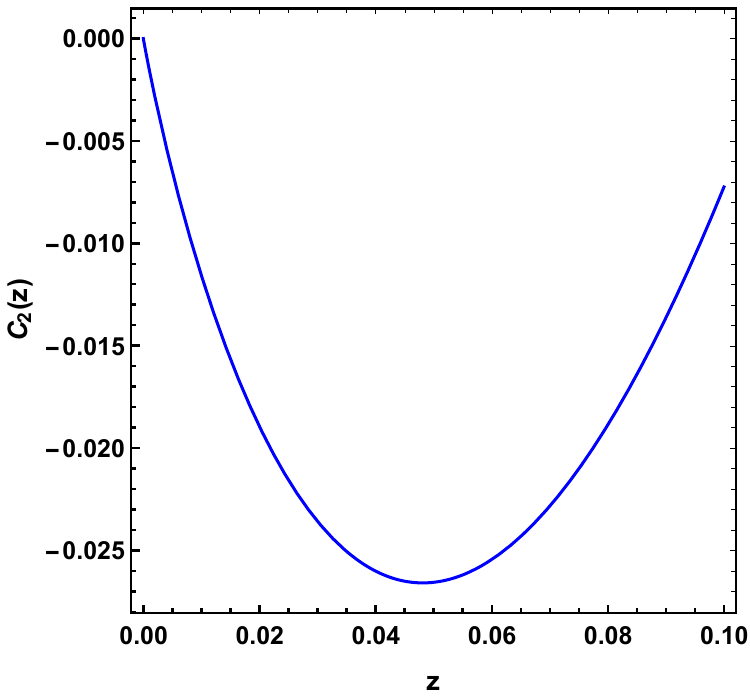}\label{figure C2}}
		\subfigure[$\ \text{C}_3(z)$]{\includegraphics[width=4.85cm]{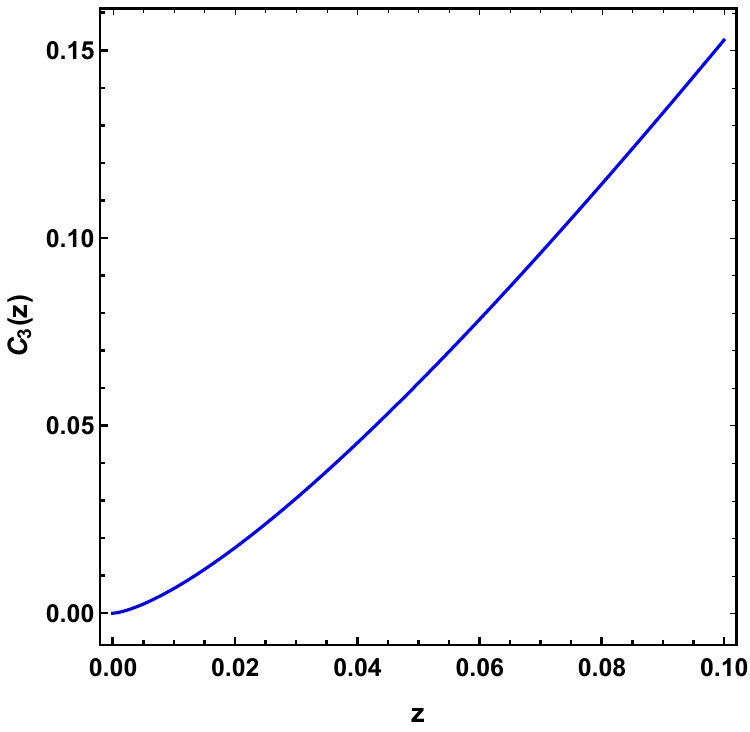}\label{figure C3}}
	\end{center}
	\caption{The numerical results of $\text{C}_1(z)$, $\text{C}_2(z)$, and $\text{C}_3(z)$ with respect to $z$.}
	\label{figure C}
\end{figure}

In this appendix, we give the numerical results on the several defined functions $\text{C}_1(z)$,  $\text{C}_2(z)$ and  $\text{C}_3(z)$ in the analytical expressions of the gravitational deflection angle. The detailed expressions of these functions are:
\begin{subequations}
    \begin{eqnarray}
        &&\text{C}_1(z)=z^2\int_0^{\pi } \csc ^2\phi  \, _2F_1\left(\frac{2}{3},1;\frac{5}{3};-z^3 \csc^3 \phi \right) \, \rm{d}\phi, \ \ \ \ \ \ \ \ 
        \\
        &&\text{C}_2(z)=z\int_0^{\pi } \arctan \frac{2 \sqrt{ z\csc \phi }-1}{\sqrt{3}} \, \rm{d}\phi, \ \ \ \ \ \ \ \ 
        \\
        &&\text{C}_3(z)=z\int_0^{\pi } \mathrm{arctanh} \frac{3 \sqrt{z \csc \phi }}{2+2 z \csc \phi +\sqrt{z \csc \phi }}\, \rm{d}\phi. \ \ \ \ \ \ \ \ 
    \end{eqnarray}
\end{subequations}

The numerical results of $\text{C}_1(z)$, $\text{C}_2(z)$, and $\text{C}_3(z)$ with respect to $z$ are shown Fig. \ref{figure C}. As can be seen from the figure, $\text{C}_1(z)$ also increases and presents a linear relationship with the increase of $z$, while $\text{C}_2(z)$ first increases and then decreases, resulting in a local minimum, and $\text{C}_3(z)$ also increases with $z$ but has a non-linear behavior.

\acknowledgments

The authors are grateful to Peng Wang, Aoyun He, Yang Huang, and Yadong Xue for useful discussions. The authors would like to thank the anonymous referee for helpful comments and suggestions, which helped to improve the quality of this paper. This work is supported by the National Natural Science Foundation of China (Grant No. 12147207, No. 12175212, and No. 12275184), the “zhitongche” program for doctors from Chongqing Science and Technology Committee (Grant No. CSTB2022BSXM-JCX0100), the Natural Science Foundation of Chongqing Municipality (Grant No. CSTB2022NSCQ-MSX0932), the Scientific and Technological Research Program of Chongqing Municipal Education Commission (Grant No. KJQN202201126).

\bibliographystyle{unsrt}

\end{document}